\documentclass{bmcart}

\usepackage{amsthm,amsmath,amsfonts,amssymb}
\usepackage[title]{appendix}

\usepackage{color}
\definecolor{hilite}{rgb}{0,0,0}
\providecommand{\chg}[1]{\textcolor{hilite}{#1}}

\RequirePackage{mathtools}
\RequirePackage{natbib}
\usepackage{hyperref}
\usepackage[none]{hyphenat}
\usepackage{nicefrac}
\usepackage{pgfplots}
\usetikzlibrary{arrows, decorations.markings, plotmarks}
\input{Qcircuit.tex}
\usepackage{bbold}
\usepackage{color}
\usepackage{stmaryrd}
\usepackage{calc}
\usepackage{verbatim}
\usepackage{tikz}
\usetikzlibrary{calc}

\newsavebox\CBox
\newcommand\hcancel[2][0.5pt]{%
  \ifmmode\sbox\CBox{$#2$}\else\sbox\CBox{#2}\fi%
  \makebox[0pt][l]{\usebox\CBox}%
  \rule[0.5\ht\CBox-#1/2]{\wd\CBox}{#1}}

\providecommand{\ket}[1]{\left \vert #1 \right \rangle}
\providecommand{\bra}[1]{\left \langle #1 \right \vert}
\providecommand{\braket}[2]{\left \langle #1 \left \vert #2 \right. \right \rangle}

\providecommand{\elem}[3]{\left \langle #1 \left \vert \vphantom{#1#2#3} #2 \right \vert #3 \right \rangle}

\providecommand{\ketbra}[2]{\ket{#1} \! \bra{#2}}
\providecommand{\proj}[1]{\ketbra{#1}{#1}}

\providecommand{\abs}[1]{\left \vert #1 \right \vert}
\providecommand{\set}[1]{\left \lbrace #1 \right \rbrace}

\providecommand{\id}{\hat{\mathbb{1}}}
\providecommand{\com}[2]{\left[#1,\,#2 \right]}
\providecommand{\acom}[2]{\left \lbrace #1,\,#2 \right \rbrace}
\providecommand{\diss}[2]{\mathcal{D}\left[ #1 \right]\left( #2 \right)}
\providecommand{\meas}[2]{\mathcal{M}\left[ #1 \right]\left( #2 \right)}
\providecommand{\lindtwo}[2]{ #1 #2 #1^{\dagger} - \dfrac{1}{2} \left \lbrace #1^{\dagger} #1,\,#2 \right \rbrace }

\providecommand{\tr}{\mathrm{tr}}
\providecommand{\trace}[1]{\mathrm{tr} \left( #1 \right)}

\providecommand{\set}[1]{\left \lbrace #1 \right \rbrace}
\renewcommand{\Im}{\textrm{Im}}
\renewcommand{\Re}{\textrm{Re}}
\providecommand{\lind}{\mathcal{L}}
\providecommand{\norm}[2]{\left \vert \left \vert #2 \right \vert \right \vert_{#1}}

\providecommand{\ct}{^{\dagger}}

\providecommand{\norm}[1]{\left\Vert #1 \right\Vert} 

\makeatletter
\newcommand{\pushright}[1]{\ifmeasuring@#1\else\omit\hfill$\displaystyle#1$\fi\ignorespaces}
\newcommand{\pushleft}[1]{\ifmeasuring@#1\else\omit$\displaystyle#1$\hfill\fi\ignorespaces}
\makeatother
\newlength\figureheight
\newlength\figurewidth
\startlocaldefs
\endlocaldefs

\begin{document}

\begin{frontmatter}

\begin{fmbox}
\dochead{Research}

\title{Multi-Qubit Joint Measurements in Circuit QED: Stochastic Master Equation Analysis}

\author[
   addressref={aff1},                   
   corref={aff1},                       
   email={criger@physik.rwth-aachen.de}   
]{\inits{DB}\fnm{Ben} \snm{Criger}}
\author[
   addressref={aff1},                   
   email={ciani@physik.rwth-aachen.de}   
]{\inits{A}\fnm{Alessandro} \snm{Ciani}}
\author[
   addressref={aff1,aff2},
   email={d.divincenzo@fz-juelich.de}
]{\inits{DP}\fnm{David P} \snm{DiVincenzo}}

\address[id=aff1]{
  \orgname{JARA Institut F\"{u}r Quanteninformation, RWTH}, 
  \street{Otto-Blumenthalstra{\ss}e 20},                     %
  \postcode{52074}                                
  \city{Aachen},                              
  \cny{Germany}                                    
}
\address[id=aff2]{%
  \orgname{Forschungszentrum J\"{u}lich},
  \postcode{52425}
  \city{J\"{u}lich},
  \cny{Germany}
}

\begin{artnotes}
\end{artnotes}

\end{fmbox}

\begin{abstractbox}

\begin{abstract}
We derive a family of stochastic master equations describing homodyne measurement of multi-qubit diagonal observables in circuit quantum electrodynamics. \chg{In the regime where qubit decay can be neglected,} our approach replaces the polaron-like transformation of previous work, which required a lengthy calculation for the physically interesting case of three qubits and two resonator modes.  The technique introduced here makes this calculation straightforward and manifestly correct. Using this technique, we are able to show that registers larger than one qubit evolve under a non-Markovian master equation. We perform numerical simulations of the three-qubit, two-mode case from previous work, obtaining an average post-measurement state fidelity of $\sim$ 94\% \chg{, limited by measurement-induced decoherence and dephasing}. 
\end{abstract}


\end{abstractbox}

\end{frontmatter}

\section{Introduction}
Circuit QED provides a promising avenue for the realization of quantum algorithms, with recent experiments showing increases in both coherence time and precision of control \cite{SingleQubitGates, TwoQubitGates, SingleShotMeasurement}. Quantum algorithms are thought to require error correction as a prerequisite \cite{NielsenChuang}, and quantum error correction requires non-demolition measurement of joint operators, most often Pauli operators of low weight \cite{ToricCode, BaconShorCodes, ColourCodes, SparseGraphCodes}. This can be accomplished using an ancilla register which is prepared in a specific state, interacts with the encoded state, and is then measured (possibly destructively) \cite{ShorEC, SteaneEC, KnillEC, DKLP}. In circuit QED, ancilla measurement has been accomplished by coupling the qubit to photons passing through a resonator, and observing the accrued phase using homodyne detection \cite{SingleQubitHomodyneMeasurement}. 

Recent work has begun to consider \emph{direct} joint measurements in circuit QED, in which all qubits in the support of the measured operator are coupled to one or more \chg{internal} resonator modes, using homodyne detection to observe an output mode, requiring no ancilla qubit. Difficulty in calculating the reduced qubit dynamics has restricted previous analysis of direct measurement schemes to systems containing two \chg{\cite{TunableJointMeasurement, UndoingDephasing, RemoteEntanglement}} or three \cite{SolgunDiV, TBDiV} qubits. In this paper, we simplify this calculation, deriving reduced qubit dynamics for an arbitrary number of qubits and resonator modes. We then use the resulting stochastic master equation to extend the analysis of the three-qubit, two-mode scheme presented in \cite{SolgunDiV, TBDiV}.

The rest of this paper is organized as follows. We write the multi-qubit, multi-mode Lindbladian in Section \ref{sec:params}, and incorporate it into a stochastic master equation corresponding to homodyne measurement of the output mode. In Sections \ref{sec:cav_eoms} and \ref{sec:reg_eoms} we determine the reduced equations of motion for the resonator and register states, respectively. Using these equations, we proceed to simulate multi-qubit measurement dynamics in section \ref{sec:sims}. We discuss what can be done to increase post-measurement state fidelity and conclude in section \ref{sec:conc}.
\section{Parameters}
\label{sec:params}
We begin with a set of qubits $Q$, called the register, and a set of internal resonator modes $C$. For convenience, we denote as $B$ the set of $0/1$ assignments to the register (comprising $n=\abs{Q}$ bits). We consider a scenario in which the coupling between the internal modes and the input/output mode is described by the relation
\begin{equation}
a_{\mathrm{out}} = \sum_{k \in C} \sqrt{\kappa_k}a_k. \label{eq:a_out}
\end{equation}
\chg{(Recall that the discrepancy in units is explained by comparing the waveguide commutation relation, $\com{a_{\mathrm{out}}(\omega)}{a \ct_{\mathrm{out}}(\omega')} = \delta(\omega - \omega')$, and the cavity commutation relation $\com{a_k}{a \ct_{k'}} = \delta_{k,k'}$.)}
 A single measurement tone is used, and we describe the dynamics in a frame rotating at the carrier frequency of this tone. 

\chg{We consider a model Hamiltonian for the coupled-qubit system that describes the essential aspects of the dispersive-coupling regime, that is, when the qubit frequencies are far detuned from the cavity frequencies:}
\begin{flalign}
H_{\textrm{Disp}} &= \sum_{k \in C}\Delta_{k} a_k^{\dagger}a_k + \sum_{l \in Q} \left(\dfrac{\Omega_l}{2}+ \sum_{k \in C} \chi_{k,l} \right) \sigma_{z,l} + \sum_{l \in Q, k \in C} \chi_{k,l}\sigma_{z,l} a_k^{\dagger}a_k \nonumber \\
& + \epsilon(t) \sum_{k \in C} \sqrt{\kappa_k } \left(a_k + a_k^{\dagger}\right)\label{dispeq}.
\end{flalign}
Here, $\hbar=1$, $\Delta_k \triangleq \omega_k - \omega$ is the difference between the $k$th resonator mode frequency and the measurement tone frequency $\omega$, $a_k$ $\left(a^{\dagger}_k\right)$ is the lowering (raising) operator on the $k$th resonator mode, $\chi_{k,l}$ is the coupling frequency between the $k$th resonator mode and the $l$th qubit, $\Omega_l + \sum_{k \in C} \chi_{k,l} $ is the Lamb-shifted qubit frequency (with $\Omega_l$ being the bare frequency), and $\epsilon(t)$ is the time-dependent measurement tone amplitude.

\chg{We note that if we use a Schrieffer-Wolff analysis to derive the dispersive-coupling Hamiltonian from an underlying multi-qubit Jaynes-Cummings model \cite{ExploringTheQuantum}, additional terms appear, which describe qubit-qubit and resonator-resonator couplings\cite{PhysRevA.69.062320, ModeModeCoupling}.  In Appendix~\ref{appa} we  give a full derivation of this expression.  We give arguments for why these terms can be neglected (within a rotating-wave approximation) or incorporated into parity measurement schemes with straightforward modifications of the analysis described below. It is also interesting to note that, with a more general starting point provided by circuit Hamiltonians  \cite{PhysRevB.78.104508}, couplings more general than the Jaynes-Cummings form appear, and some qubit-qubit and mode-mode coupling terms can be arranged to cancel, as explored in \cite{PhysRevB.78.104508}. We proceed with the model Hamiltonian Eq. (\ref{dispeq}), as it permits a full and clear exploration of all the issues connected with parity measurement, without the inessential complicating features introduced by the additional coupling terms.}

We assume that decoherence can be described using a Lindblad master equation, consisting of terms of the form $\diss{L}{\rho} = \lindtwo{L}{\rho}$. The noise sources we include in the model, as in previous work \chg{\cite{SingleQubitHomodyneMeasurement, TunableJointMeasurement, UndoingDephasing, TBDiV, RemoteEntanglement}} are resonator photon loss, intrinsic dephasing/amplitude damping of the qubit states, and the Purcell effect \cite{ExploringTheQuantum}:
\begin{flalign}
\label{eq:me}
\dot{\rho} = \mathcal{L}(\rho) &= -i\com{H_{\textrm{Disp}}}{\rho} + \diss{\sum_{k \in C} \sqrt{\kappa_k} a_k}{\rho} + \dfrac{1}{2} \sum_{l \in Q} \gamma_{z,l} \diss{\sigma_{z,l}}{\rho} \nonumber \\
& + \sum_{l \in Q} \gamma_{-,l}  \diss{\sigma_{-,l}}{\rho} + \diss{\sum_{k \in C, l \in Q} \sqrt{\kappa_k} \lambda_{k,l} \sigma_{-,l} }{\rho},
\end{flalign} 
where $\kappa_k$ is the photon loss rate in the $k$th resonator mode, $\gamma_{z,l}$ is the intrinsic dephasing rate for the $l$th qubit, $\gamma_{-,l}$ is the intrinsic amplitude damping rate for the $l$th qubit, and $\lambda_{k,l}$ is \chg{an effective} Purcell factor for the $k$th mode interacting with the $l$th qubit.

\chg{Since, as emphasized by the derivation of Appendix \ref{appa}, the factors $\lambda_{k,l}$  can have either sign, it is perfectly possible to arrange these factors so that the $k$ sum in the final term of Eq. (\ref{eq:me}) is zero for every $l$.  In other words, an effective Purcell filter \cite{PurcellFiltering} can be created by taking advantage of flexibility provided by the multi-mode structure.  Given the ongoing advances in qubit coherence, we believe it is also reasonable to ignore intrinsic qubit damping, i.e., we can set $\gamma_{-,l}=0$.  Thus, from this point onward, we will ignore qubit damping effects (but we will retain qubit dephasing terms). } 

To model the evolution of the state $\rho$ and measurement record $j$ under homodyne measurement, we use the stochastic master equation \cite{QuantumQuadratureMeasurements}:
\begin{flalign}
\label{eq:sme}
&d\rho = \lind(\rho)dt + \sqrt{\eta} \meas{e^{-i\phi} a_{\mathrm{out}}}{\rho}dW \\ 
&j(t)dt = \sqrt{\eta} \left \langle e^{-i\phi} a_{\mathrm{out}} + e^{i\phi} a_{\mathrm{out}}\ct \right \rangle dt + dW \label{eq:photocurrent} \\ 
&\textrm{ where } \meas{c}{\rho} = c \rho + \rho c^{\dagger} - \tr \left( c \rho + \rho c^{\dagger} \right)\rho \label{eq:m_c_rho}
\end{flalign}
Here, $\eta \in [0,1]$ is the quantum efficiency of the homodyne measurement, $\phi$ is the homodyne phase (which we set to $0$, corresponding to measurement of the real part of the operator $a_{\mathrm{out}}$) and $dW$ is a Wiener increment (a normal variate with mean 0 and variance $dt$)\cite{NumericalSolutionSDEs}. 

\chg{As proved in Appendix \ref{appb}, a} family of solutions to Eq. (\ref{eq:me}) or Eq. (\ref{eq:sme}) can be expressed using pointer states \cite{PRepresentation, PointerStates}:
\begin{equation}
\label{eq:state_ansatz}
\rho = \sum_{i,j \in B} \rho_{i,j}(t)  \ketbra{i}{j} \otimes \bigotimes_{k \in C} \ketbra{\alpha_{k,i}(t)}{\alpha_{k,j}(t)}.
\end{equation}
Here, $\ket{\alpha_{k,j}(t)}$ is a coherent state (an eigenstate of the lowering operator $a_k$ with eigenvalue $\alpha_{k,j}(t)$) corresponding to the bitstring $j$. This simplifies the numerical solution of the (deterministic or stochastic) master equation in the event that the initial state is coherent (the vacuum is such a state, with $\alpha_{k,j} = 0$ $\forall k,j$), by requiring only a fixed number of coherent state trajectories to be calculated, rather than the full, infinite-dimensional resonator state.

In the following sections, we simplify the equations of motion further, by deriving the equations of motion for the coherent state amplitudes $\set{\alpha_{k,j}}$, and incorporating these into the equation of motion for $\rho_Q$, the register reduced state.
\section{Resonator Equations of Motion}
\label{sec:cav_eoms}
Equations of motion for the resonator mode lowering operator $a_k$ can be derived using input-output theory \cite{QuantumOptics}:
\begin{flalign}
\dot{a}_k(t) & = i\com{H_{\textrm{Disp}}}{a_k(t)} - \sum_{k'} \frac{\sqrt{\kappa_k \kappa_{k'}}}{2}  a_{k'}(t)\\
&= -i\Delta_k a_k(t) - i\sum_{l} \chi_{k,l} \sigma_{z,l}a_k(t) - i\sqrt{\kappa_k}\epsilon(t) - \sum_{k'} \dfrac{ \sqrt{\kappa_k \kappa_{k'} }}{2} a_{k'}(t)
\end{flalign}
The corresponding equation for $\dot{\alpha}_{k,j}(t)$ can be found as in \cite{TBDiV}:
\begin{flalign}
\dot{\alpha}_{k,j}(t) = & -i\Delta_k \alpha_{k,j}(t) - i\sum_{l} \chi_{k,l} (-1)^{j_l} \alpha_{k,j}(t) - i\sqrt{\kappa_k}\epsilon(t)  \nonumber \\
& - \sum_{k'} \dfrac{ \sqrt{\kappa_k \kappa_{k'} }}{2} \alpha_{k',j}(t)
\label{eq:cavity_eom}
\end{flalign}
where $j_l$ is the value of the $l$th bit of $j$. The amplitude of the output of the system is (see Eq. (\ref{eq:a_out})):
\begin{equation}
\alpha_{\mathrm{out}} = \sum_{k \in C} \sqrt{\kappa_k}\alpha_{k,j}(t)
\end{equation}
In the following subsection, we place this system of linear, first-order \chg{ordinary differential equations (ODEs)} into a canonical form which can be used for further analysis.
\subsection{State-Space Representation}
Systems of first-order linear ODEs with time-invariant coefficients can be represented using a vector $\vec{x}$, called the state, a vector $\vec{u}(t)$, called the input, and a vector $\vec{y}(t)$, called the output. They can be written in a standard form:
\begin{flalign}
\dot{\vec{x}}(t) &= A\vec{x}(t) + B\vec{u}(t) \nonumber \\
\vec{y}(t)       &= C\vec{x}(t) + D\vec{u}(t).
\end{flalign} 
This is the popular state-space representation of linear time-invariant (LTI) systems \cite{LTISystems, NetworkAnalysis}.

For the evolution described in Eq. (\ref{eq:cavity_eom}), and fixing the register state to a specific bitstring $j$, these matrices can be written explicitly:
\begin{flalign}
&A_{k, k'} = -i \left( \Delta_k + \sum_{l \in Q}(-1)^{j_l} \chi_{k,l} \right) \delta_{k, k'} -  \dfrac{\sqrt{\kappa_k \kappa_{k'}}}{2} \nonumber \\
&B_{k,k'}= -i\sqrt{\kappa_k} \delta_{0,k'} \nonumber \\
&C_{k,k'}= \sqrt{\kappa_{k'}} \delta_{k,0} \nonumber \\
&D_{k,k'}= 0 \label{eq:abcd_matrices}
\end{flalign}
The response to a given input can be calculated in the Laplace domain using the transfer matrix $G(s)$:
\begin{equation}
Y(s) = G(s)U(s); \quad G(s) = C(s\id - A)^{-1}B + D \label{eq:transfer-matrix}
\end{equation}
In the following subsection, we use these matrices to tailor the values of the resonator parameters in order to perform a specific measurement.
\subsection{Steady States}
In general, the resonator system will exhibit a different response to a given input $\epsilon(t)$ for every distinct state of the register $\ket{j}$, $j \in B$. This is not useful for measurements of joint degrees of freedom, which should not distinguish between given subsets of $B$. Consider, as a first example, measurements of the Hamming weight $h(j)$ (the number of qubits in the ground state). Given that the system in Eq. (\ref{eq:abcd_matrices}) only depends on $j$ through the term $\sum_{l \in Q}(-1)^{j_l} \chi_{k,l}$ in the $A$ matrix, systems with identical Hamming weights have identical $(A,\,B,\,C,\,D)$ matrices if all $\chi_{k,l}$ are equal to a single constant $\chi$:
\begin{equation}
\sum_{l \in Q}(-1)^{j_l} \chi = \left(\abs{Q} - 2h(j)\right)\chi.
\end{equation}
To perform a parity measurement, it is furthermore necessary for the responses to be close for all even $h$ and all odd $h$, with the even and odd responses being different, to ensure distinguishability. To determine whether this is possible for the three-qubit case, we analyze the difference between real transfer matrices corresponding to $h=0$ and $h=2$:
\begin{flalign}
& G_{h = 0}(s) - G_{h = 2}(s) =
\dfrac{1}{d(s)}\begin{bmatrix}
 a(s) & -b(s) \\
 b(s) & a(s) \\
\end{bmatrix}, 
\end{flalign}
where $a(s)$, $b(s)$, and $d(s)$ are functions of $\set{\kappa_k}$, $\set{\Delta_k}$ too lengthy to include here. Seeing that this matrix is full-rank, we determine that there is no complex driving function $\epsilon(t)$ which results in exactly equal outputs at all time. 

To obtain an approximate three-qubit parity measurement for large measurement times, it is sufficient to set the steady-state responses equal for $h=0$ and $h=2$ (equality of the $h=1$ and $h=3$ responses follows by symmetry). In order to accomplish this, we note that:
\begin{equation}
\vec{y}_{ss} = G(0)\vec{u}_{ss} = \left( -C A^{-1} B + D \right) \vec{u}_{ss}
\end{equation}

$A$ can be expressed as the sum of a diagonal matrix and a rank-one matrix. This allows us to invert it analytically, using the Sherman-Morrison formula \cite{ShermanMorrison}:
\begin{flalign}
&A = A' + \vec{v}\vec{v}^{\intercal}\nonumber \\
&\textrm{where }A'_{k,k'} = -i \left( \Delta_k + \sum_{l \in Q}(-1)^{j_l} \chi_{k,l} \right) \delta_{k, k'} \nonumber \\
&\textrm{and } \vec{v}_k = i\sqrt{\dfrac{\kappa_k}{2}}\\
&\therefore A^{-1} = A'^{-1} - \dfrac{A'^{-1}\vec{v}\vec{v}^{\intercal}A'^{-1}}{1 + \vec{v}^{\intercal}A'^{-1}\vec{v}}\\ 
& A^{-1}_{k,k'} = i\dfrac{\delta_{k,k'}}{\tilde{\Delta}_k} - \dfrac{\sqrt{\kappa_{k} \kappa_{k'}}}{2 \tilde{\Delta}_{k} \tilde{\Delta}_{k'} \left(1 - i\sum_{k''} \dfrac{\kappa_{k''}}{2\tilde{\Delta}_{k''}}\right)}
\end{flalign}
where $\tilde{\Delta}_k = \Delta_k + (\abs{Q} - 2h(j))\chi$.

The output corresponding to the steady state of the resonator system is:
\begin{flalign}
\alpha_{\mathrm{out} \, ss}= \frac{-i\sum_{k} \frac{\kappa_{k}}{\tilde{\Delta}_{k}}}{i + \frac{1}{2}\sum_{k} \frac{\kappa_{k}}{\tilde{\Delta}_{k}}} \epsilon_{ss}
\end{flalign}
To match steady states in the $\abs{Q}=3$ parity measurement from \cite{TBDiV}, we impose the following condition on the detunings of the two resonator modes ($\Delta_0$, $\Delta_1$):
\begin{flalign}
\label{eq:deltas}
\frac{\Delta_0}{\chi} = \sqrt{3\frac{\kappa_0}{\kappa_1}}, \,\, \frac{\Delta_1}{\chi} = -\sqrt{3\frac{\kappa_1}{\kappa_0}}
\end{flalign}
In order to constrain $\kappa$, we maximize the difference in $\Re(\alpha_{\mathrm{out}})$ between the $h=0$ and the $h=1$ cases. This is achieved when $\kappa_0 = \kappa_1 = 2\chi$.
\section{Register Equations of Motion}
\label{sec:reg_eoms}
The most economical equation for simulating the desired homodyne measurement is a reduced master equation acting only on the qubit register. Obtaining and solving such an equation has been the strategy of previous works \chg{\cite{SingleQubitHomodyneMeasurement, TunableJointMeasurement, UndoingDephasing, TBDiV, RemoteEntanglement}}. Here, we provide a simplified derivation of the reduced master equation. We begin by expressing the register state using a partial trace acting on the state ansatz given in Eq. (\ref{eq:state_ansatz}):
\begin{flalign}
\rho_Q \triangleq \tr_C(\rho) = \sum_{i,j \in B} \rho_{i,j} \prod_{k \in C} \braket{\alpha_{k,j}}{\alpha_{k,i}} \ketbra{i}{j}, \,\, \rho_{Q \, i,j} =  \rho_{i,j} \prod_{k \in C} \braket{\alpha_{k,j}}{\alpha_{k,i}}.
\label{eq:resonator_trace}
\end{flalign}
In order to determine the dynamics of this reduced state, we use the integral representation of the partial trace over the harmonic oscillator \cite{CoherentStates}:
\begin{flalign}
\tr_C(\rho) = \dfrac{1}{\pi^{\abs{C}}} \left[ \int_{\vec{\alpha} \in \mathbb{C}^{\abs{C}}} \elem{\vec{\alpha}}{\rho}{\vec{\alpha}} d^2 \vec{\alpha} \right] 
\end{flalign}
Being a dummy variable, the vector of amplitudes $\vec{\alpha}$ in the integral above does not depend on time or on the Wiener increment, resulting in a simple expression for the reduced master equation:
\begin{flalign}
d \rho_Q = \dfrac{1}{\pi^{\abs{C}}} \int_{\vec{\alpha} \in \mathbb{C}^{\abs{C}}} \elem{\vec{\alpha}}{\mathcal{L}(\rho)dt + \meas{a_{\mathrm{out}}}{\rho}dW}{\vec{\alpha}} d^2 \vec{\alpha}
\end{flalign}
We evaluate the deterministic and stochastic terms separately in the following subsections.
\subsection{Deterministic}
Before calculating the partial trace of the deterministic Lindbladian $\mathcal{L}$, we note that, for an operator $R$ supported only on the resonator Hilbert spaces, the partial trace acting on a commutator or dissipator annihilates it:
\begin{flalign}
&\tr_{C}\left( \com{R}{\rho} \right) = \sum_{i,j \in B} \rho_{i,j} \ketbra{i}{j} \times \tr\left( \com{R}{\bigotimes_{k \in C} \ketbra{\alpha_{k,i}}{\alpha_{k,j}}} \right) = 0 \label{eq:res-com-elim}\\
&\tr_{C}\left( \diss{R}{\rho} \right) = \sum_{i,j \in B} \rho_{i,j} \ketbra{i}{j} \times \tr\left( \diss{R}{\bigotimes_{k \in C} \ketbra{\alpha_{k,i}}{\alpha_{k,j}}} \right) = 0 \label{eq:res-diss-elim}
\end{flalign}
Also, for any operator $Q$ supported only on the register, the partial trace acting on a commutator or dissipator is expressed in terms of inner products of coherent states, similar to the partial trace of $\rho$ in Eq. (\ref{eq:resonator_trace}):
\begin{flalign}
\tr_{C}\left( \com{Q}{\rho} \right) &= \sum_{i,j \in B} \rho_{i,j} \com{Q}{\ketbra{i}{j}} \times \tr\left( \bigotimes_{k \in C} \ketbra{\alpha_{k,i}}{\alpha_{k,j}} \right)  \nonumber \\
&= \sum_{i,j \in B} \rho_{i,j} \com{Q}{\ketbra{i}{j}} \times \prod_{k \in C} \braket{\alpha_{k,j}}{\alpha_{k,i}}  \nonumber \\
&= \sum_{i,j \in B} \rho_{Q \, i,j} \com{Q}{\ketbra{i}{j}} = \com{Q}{\rho_Q} \label{eq:reg-com-simp} \\
\tr_{C}\left( \diss{Q}{\rho} \right) &= \sum_{i,j \in B} \rho_{i,j} \diss{Q}{\ketbra{i}{j}} \times \tr \left( \bigotimes_{k \in C} \ketbra{\alpha_{k,i}}{\alpha_{k,j}} \right)  \nonumber \\
&= \sum_{i,j \in B} \rho_{i,j} \diss{Q}{\ketbra{i}{j}} \times \prod_{k \in C} \braket{\alpha_{j,k}}{\alpha_{i,k}}  \nonumber \\
&= \sum_{i,j \in B} \rho_{Q\, i,j} \diss{Q}{\ketbra{i}{j}} = \diss{Q}{\rho_Q} \label{eq:reg-diss-simp}
\end{flalign}
The only partial trace which does not benefit from these simplifications is $\tr_C \left(\com{\sigma_{z,l}a_{k'}^{\dagger}a_{k'}}{\rho}\right)$. To write it succinctly, we introduce modified density matrices $\rho(O_{k'})$ and $\rho_Q(o_{k'})$, where $O_{k'}$ is an operator on the $k$th resonator space, and $o_{k'}$ is a scalar:
\begin{flalign}
&\rho(O_{k'}) \triangleq \sum_{i,j \in B} \rho_{i,j}  \ketbra{i}{j} \otimes \bigotimes_{k \in C, k < k'} \ketbra{\alpha_{k,i}}{\alpha_{j,k}} \otimes O_{k'} \otimes \bigotimes_{k \in C, k > k'} \ketbra{\alpha_{i,k}}{\alpha_{j,k}} \label{eq:rho_op_func}\\
&\rho_Q(o_{k'}) \triangleq \sum_{i,j \in B} \rho_{i,j}  \ketbra{i}{j} \times \prod_{k \in C, k < k'} \braket{\alpha_{k,j}}{\alpha_{k,i}} \times o_{k'} \times \prod_{k \in C, k > k'} \braket{\alpha_{k,j}}{\alpha_{k,i}}, \label{eq:rho_scal_func}
\end{flalign}
noting that $O_{k'}$ and $o_{k'}$ may also depend on the register basis state $\ket{j}$, or on other variables. 

The commutator and its partial trace can now be easily expressed:
\begin{flalign}
&\com{\sigma_{z,l}a_{k'}^{\dagger}a_{k'}}{\rho} = \rho \left((-1)^{i_l} a_{k'}^{\dagger} a_{k'} \ketbra{\alpha_{k',i}}{\alpha_{k',j}} -  \ketbra{\alpha_{k',i}}{\alpha_{k',j}} (-1)^{j_l} a_{k'}^{\dagger} a_{k'} \right) \\
&\tr_{C}\left(\com{\sigma_{z,l} a_{k'}^{\dagger} a_{k'} }{\rho}\right) \nonumber \\
& = \rho_Q \left( \dfrac{1}{\pi} \int_{\alpha_{k'}\in \mathbb{C}} \bra{\alpha_{k'}} \left((-1)^{i_l} a_{k'}^{\dagger} a_{k'} \ketbra{\alpha_{k',i}}{\alpha_{k',j}} \right. \right.  \nonumber \\
& \qquad \qquad \qquad \qquad  \qquad \qquad \left. \left. -  \ketbra{\alpha_{k',i}}{\alpha_{k',j}} (-1)^{j_l} a_{k'}^{\dagger} a_{k'} \right) \ket{\alpha_{k'}} d^2 \alpha_{k'} \vphantom{\frac{1}{\pi}} \right)
\label{eq:integral_equation}
\end{flalign}
Permuting terms, we write the integral in Eq. (\ref{eq:integral_equation}) as a matrix element of the state of the $k'$th resonator mode:
\begin{flalign}
&\dfrac{1}{\pi}\int_{\alpha_{k'}\in \mathbb{C}} \bra{\alpha_{k'}} \left((-1)^{i_l} a_{k'}^{\dagger} a_{k'} \ketbra{\alpha_{i,k'}}{\alpha_{j,k'}} -  \ketbra{\alpha_{i,k'}}{\alpha_{j,k'}} (-1)^{j_l} a_{k'}^{\dagger} a_{k'} \right) \ket{\alpha_{k'}} d^2 \alpha_{k'} \nonumber \\
&= \elem{\alpha_{j,k'}}{\left[ \dfrac{1}{\pi}\int_{\alpha_{k'}\in \mathbb{C}} \left((-1)^{i_l} \bar{\alpha}_{k'}\alpha_{i,k'} -  (-1)^{j_l}\bar{\alpha}_{j,k'}\alpha_{k'} \right) \proj{\alpha_{k'}} d^2 \alpha_{k'}\right]}{\alpha_{i,k'}} 
\label{eq:permuted_integral_equation}
\end{flalign}
We use the resolution of polynomials in the raising and lowering operators \cite[Section 2.9.3]{CoherentStates}:
\begin{equation}
\sum_{m,n}c_{m,n}a^m a^{\dagger n} = \dfrac{1}{\pi}\int_{\alpha \in \mathbb{C}} c_{m,n}\alpha^m \bar{\alpha}^n \proj{\alpha} d^2 \alpha
\end{equation}
The integral in Eq. (\ref{eq:permuted_integral_equation}) reduces to:
\begin{flalign}
&\elem{\alpha_{j,k'}}{\left((-1)^{i_l} a^{\dagger}_{k'}\alpha_{i,k'} - (-1)^{j_l}\bar{\alpha}_{j,k'}a_{k'} \right)}{\alpha_{i,k'}} \nonumber \\
&= \left( (-1)^{i_l} - (-1)^{j_l} \right) \bar{\alpha}_{j,k'}\alpha_{i,k'} \braket{\alpha_{j,k'}}{\alpha_{i,k'}}
\end{flalign}
We substitute back in:
\begin{flalign}
\label{eq:coupling-lindbladian}
&\tr_{C} \left( \com{ \sigma_{z, l} a_{k'}^{\dagger}a_{k'}}{\rho}\right) \nonumber \\
& = \sum_{i,j \in B} \rho_{i,j}  \ketbra{i}{j} \prod_{k \in C} \braket{\alpha_{j,k}}{\alpha_{i,k}} \left( (-1)^{i_l} - (-1)^{j_l} \right) \bar{\alpha}_{j,k'}\alpha_{i,k'} \nonumber \\
& = \tilde{P}_{k'} \circ \com{\sigma_{z,l}}{\rho_Q}, \\
\label{eq:P_k}
& \textrm{where } \left[\tilde{P}_{k}\right]_{i,j} \triangleq \bar{\alpha}_{j,k} \alpha_{i,k}.
\end{flalign}
Here, $\left(A \circ B \right)_{i,j} \triangleq A_{i,j} B_{i,j}$ defines the elementwise (or Hadamard) matrix product.

We write the unconditional master equation by transforming the qubit-only Lindbladian, and adding on a term which accounts for the qubit-resonator coupling (see Eq. \ref{eq:me}, \ref{eq:res-com-elim} -- \ref{eq:reg-diss-simp}, and \ref{eq:coupling-lindbladian}):
\begin{flalign}
\label{eq:unrot-frame-eqn}
\dot{\rho}_Q = -i \com{H_Q}{\rho_Q} + \dfrac{1}{2} \sum_{l \in Q} \gamma_{z,l} \diss{\sigma_{z,l}}{\rho_Q} -i \sum_{k\in C, l \in Q} \chi_{k,l} \tilde{P}_k \circ \com{\sigma_{z,l}}{\rho_Q}
\end{flalign}
where $H_Q = \frac{1}{2} \sum_{l \in Q} \left(\Omega_l + \sum_{k \in C} \chi_{k,l} \right) \sigma_{z,l}$.

In order to simulate evolution under this master equation, it is convenient to eliminate fast-rotating terms by expressing the master equation in a frame rotating with $H_Q$. This has the effect of eliminating $H_Q$ from the Lindbladian:
\begin{flalign}
\label{eq:uncond-master-eqn}
\dot{\rho}_Q \mapsto \dfrac{1}{2} \sum_{l \in Q} \gamma_{z,l} \diss{\sigma_{z,l}}{\rho_Q} -i \sum_{k\in C, l \in Q} \chi_{k,l} \tilde{P}_k \circ \com{\sigma_{z,l}}{\rho_Q}.
\end{flalign}
\subsubsection{Markovianity}
It is interesting to note that the coupling Lindbladian in Eq. (\ref{eq:coupling-lindbladian}), though it generates a \chg{completely-positive trace-preserving} map, is non-Markovian. This is not surprising, since the Markov approximation is the result of a weak-coupling assumption, and fast quantum measurement requires strong coupling. This has been confirmed in the case of a single-qubit measurement. The coupling Lindbladian, though it can be written in explicit Lindblad form, has a  decay rate associated with the dephasing operator which is not necessarily positive \cite{SingleQubitHomodyneMeasurement}. In this section, we prove non-Markovianity of the coupling Lindbladian in the general case, and we examine the consequences of this property of the Lindbladian by numerical simulation. 

To show that the coupling Lindbladian is non-Markovian, we note that the action of a Markovian Lindbladian on a density matrix in an $N$-dimensional Hilbert space can be expressed as \cite{GenericLindbladForm, BreuerPetruccione}:
\begin{equation}
\mathcal{L} \rho = -i\com{H}{\rho} + \acom{G}{\rho} + \sum_{i,j=1}^{N^2-1} \frac{a_{i,j}}{\left \Vert F_i \right \Vert \left \Vert F_j \right \Vert} F_i \rho F_j\ct
\label{eq:generic-lindblad-form}
\end{equation}
where the coefficients $\frac{a_{i,j}}{\left \Vert F_i \right \Vert \left \Vert F_j \right \Vert}$ form a Hermitian, positive-semidefinite matrix, and the operators $F_i$ form an orthogonal basis under the Hilbert-Schmidt inner product, with $F_0=\nicefrac{\id}{\sqrt{N}}$. In the remainder of this section, we will derive a minimal set of operators $F_i$ for our coupling Lindbladian, and show that the resulting coefficient matrix is non-positive, precluding a true Lindblad representation.

In order to express the Hadamard product from Eq. (\ref{eq:coupling-lindbladian}) in the form given by Eq. (\ref{eq:generic-lindblad-form}), we use a dyadic product formula (see \cite{HadamardProducts} for further applications of Hadamard and dyadic products to Lindbladians). The term of interest is given by:
\begin{equation}
\mathcal{L}_c \rho_Q = -i \sum_{k \in C} \sum_{l \in Q} \chi_{k,l}\tilde{P}_k \circ \com{\sigma_{z,l}}{\rho_Q}
\label{eq:cpl-lind}
\end{equation}
The matrix $\tilde{P}_k$ given in Eq. (\ref{eq:P_k}) is a dyadic product, so we can rewrite the elementwise product as a conjugation by diagonal matrices:
\begin{equation}
\tilde{P}_k \circ \com{\sigma_{z,l}}{\rho_Q} = \hat{\alpha}_k \com{\sigma_{z,l}}{\rho_Q} \hat{\alpha}_k\ct
\end{equation}
where $\hat{\alpha}_{k \, i,j} = \alpha_{k,i} \delta_{i,j}$, and we have used Eq. (5) from \cite{HadamardProducts}. We can now express the action of the Lindbladian in terms of a non-orthonormal set of operators:
\begin{flalign}
\label{eq:non-orthonormal-set}
\mathcal{L}_c \rho_Q &= -i \sum_{k\in C, l\in Q} \chi_{k,l} \left( \hat{\alpha}_k \sigma_{z,l} \rho_Q \hat{\alpha}_k\ct - \hat{\alpha}_k \rho_Q \sigma_{z,l} \hat{\alpha}_k\ct \right) \nonumber \\
&= -i\sum_{k\in C} \hat{\alpha}_k \bar{\sigma}_{z,k} \rho_Q \hat{\alpha}_k\ct - \hat{\alpha}_k \rho_Q \bar{\sigma}_{z,k} \hat{\alpha}_k\ct 
\end{flalign}
where $\bar{\sigma}_{z,k} \triangleq \sum_{l \in Q} \chi_{k,l}\sigma_{z,l}$.
We now show the non-positivity of the coupling Lindbladian in the case of a single mode (dropping the index $k$). To put Eq. (\ref{eq:non-orthonormal-set}) in the form of Eq. (\ref{eq:generic-lindblad-form}), 
we decompose the operators $\hat{\alpha}$ and $\hat{\alpha}\bar{\sigma}_{z}$ in terms of $\id$, $F_1$ and $F_2$:
\begin{flalign}
&\hat{\alpha} = F_1 + \trace{\hat{\alpha}}\dfrac{\id}{N}\\
&\hat{\alpha}\bar{\sigma}_{z} = F_2 + \trace{\hat{\alpha}\bar{\sigma}_{z}}\dfrac{\id}{N} + \trace{\hat{\alpha}\bar{\sigma}_{z}F_1^{\dagger}}F_1
\end{flalign}
Substituting into Eq. (\ref{eq:non-orthonormal-set}), we can derive the elements of the coefficient matrix:
\begin{flalign}
&-i \hat{\alpha} \bar{\sigma}_{z} \rho \hat{\alpha}^{\dagger} + i \hat{\alpha} \rho \bar{\sigma}_z \hat{\alpha}^{\dagger} \nonumber \\
= & -i \left( F_2 + \trace{\hat{\alpha}\bar{\sigma}_{z}}\dfrac{\id}{N} + \trace{\hat{\alpha}\bar{\sigma}_{z}F_1^{\dagger}}F_1 \right) \rho \left( F_1\ct + \trace{\hat{\alpha}}^{\ast}\dfrac{\id}{N} \right) \nonumber\\
&  + i \left( F_1 + \trace{\hat{\alpha}}\dfrac{\id}{N} \right) \rho \left( F_2\ct + \trace{\hat{\alpha}\bar{\sigma}_{z}}^{\ast}\dfrac{\id}{N} + \trace{\hat{\alpha}\bar{\sigma}_{z}F_1^{\dagger}}^{\ast}F_1\ct \right)
\end{flalign}
Therefore,
\begin{flalign}
&\frac{a_{1,1}}{\left \Vert F_1 \right \Vert^2} = -i \left( \trace{\hat{\alpha} \bar{\sigma}_z F_1^{\dagger}} - \trace{\hat{\alpha} \bar{\sigma}_z F_1 \ct }^{\ast} \right) = 2 \Im \left( \trace{\hat{\alpha} \bar{\sigma}_z F_1 \ct} \right) \triangleq 2x \\
&\frac{a_{2,2}}{\left \Vert F_2 \right \Vert^2} = 0\\
&\frac{a_{2,1}}{\left \Vert F_1 \right \Vert \left \Vert F_2 \right \Vert} = -\frac{a_{1,2}}{\left \Vert F_1 \right \Vert \left \Vert F_2 \right \Vert} = -i 
\end{flalign}
Diagonalizing, we see that the eigenvalues are $x \pm \sqrt{x^2 + 1}$; the matrix has one negative eigenvalue, since $x$ is real. This, in turn, implies that the coupling Lindbladian is always non-Markovian for $\abs{C}=1$, as long as $\abs{Q}>1$. 

Note, however, that if only one qubit is present, the operators $F_0$, $F_1$ and $F_2$ derived above are always linearly dependent. \chg{Therefore, the above argument is inapplicable in the one-qubit case.  However, we observe that the coefficient of the dissipator term of the single-qubit pseudo-Lindblad equation can be negative in some time intervals during transient evolution, showing that the one-qubit evolution also has non-Markovian features\cite{SingleQubitHomodyneMeasurement}.} We expect that non-Markovianity will be the general case for multi-mode coupling Lindbladians, since the sum of multiple coefficient matrices with negative eigenvalues is not necessarily positive.

To see the effect of this non-Markovianity on the performance of the measurement, we introduce two parity eigenstates:
\begin{flalign}
\label{eq:post-measurement-states}
&\ket{\psi_+} = \dfrac{1}{2}\left( \ket{000} + \ket{011} + \ket{101} + \ket{110} \right), \nonumber \\ &\ket{\psi_-} = \dfrac{1}{2}\left( \ket{111} + \ket{100} + \ket{010} + \ket{001} \right)
\end{flalign}
and simulate their evolution under the deterministic master equation (Eq. (\ref{eq:uncond-master-eqn})), subject to a piecewise-quadratic input pulse, detailed in Fig. (\ref{fig:pulse}). Markovian dynamics produce a trace distance $\left(\nicefrac{1}{2} \, \tr \left(\sqrt{\left( \rho_+(t) - \rho_-(t) \right)^{\dagger} \left( \rho_+(t) - \rho_-(t) \right)} \right)\right)$ between the time-dependent states corresponding to initial states $\ket{\psi_{\pm}}$ which is monotonically decreasing \cite{NonMarkovianity}. Plotting this quantity in Fig. (\ref{fig:trace_distance}), we see that, as the pulse is turned off, there is a clear increase, indicating non-Markovian behaviour. 
\begin{figure}[h!]
\centering
%
%
%
%
\tikzstyle{every pin}=[fill=white,
draw=black,
font=\footnotesize]

\begin{tikzpicture}

\definecolor{color0}{rgb}{0.917647058823529,0.917647058823529,0.949019607843137}

\begin{axis}[
title={Piecewise-Quadratic Pulse},
xlabel={Dimensionless Time $\chi t$},
ylabel={Pulse Amplitude},
xmin=0, xmax=13.5,
ymin=0, ymax=0.5,
width=\figurewidth,
height=\figureheight,
xtick={0,2,4,6,8,10,12,14},
xticklabels={0,2,4,6,8,10,12,14},
ytick={0,0.1,0.2,0.3,0.4,0.5},
yticklabels={0,0.1,0.2,0.3,0.4,0.5},
xmajorgrids,
ymajorgrids
]
\addplot [line width=0.7000000000000001pt, black]
coordinates {
(0,0)
(0.1350013500135,0.00194837342081515)
(0.270002700027,0.00779349368326061)
(0.4050040500405,0.0175353607873364)
(0.540005400054001,0.0311739747330425)
(0.675006750067501,0.0487093355203788)
(0.810008100081001,0.0701414431493455)
(0.945009450094501,0.0954702976199425)
(1.080010800108,0.12469589893217)
(1.2150121501215,0.157818247086027)
(1.350013500135,0.194837342081515)
(1.4850148501485,0.235753183918634)
(1.620016200162,0.277486091736758)
(1.7550175501755,0.315370264531088)
(1.890018900189,0.349357690483787)
(2.0250202502025,0.379448369594856)
(2.160021600216,0.405642301864295)
(2.2950229502295,0.427939487292104)
(2.430024300243,0.446339925878282)
(2.5650256502565,0.46084361762283)
(2.70002700027,0.471450562525748)
(2.8350283502835,0.478160760587035)
(2.970029700297,0.480974211806692)
(3.1050310503105,0.481070235442364)
(3.240032400324,0.481070235442364)
(3.3750337503375,0.481070235442364)
(3.510035100351,0.481070235442364)
(3.6450364503645,0.481070235442364)
(3.780037800378,0.481070235442364)
(3.9150391503915,0.481070235442364)
(4.050040500405,0.481070235442364)
(4.1850418504185,0.481070235442364)
(4.320043200432,0.481070235442364)
(4.4550445504455,0.481070235442364)
(4.590045900459,0.481070235442364)
(4.7250472504725,0.481070235442364)
(4.860048600486,0.481070235442364)
(4.9950499504995,0.481070235442364)
(5.13005130051301,0.481070235442364)
(5.26505265052651,0.481070235442364)
(5.40005400054001,0.481070235442364)
(5.53505535055351,0.481070235442364)
(5.67005670056701,0.481070235442364)
(5.80505805058051,0.481070235442364)
(5.94005940059401,0.481070235442364)
(6.07506075060751,0.481070235442364)
(6.21006210062101,0.481070235442364)
(6.34506345063451,0.481070235442364)
(6.48006480064801,0.481070235442364)
(6.61506615066151,0.481070235442364)
(6.75006750067501,0.481070235442364)
(6.88506885068851,0.481070235442364)
(7.02007020070201,0.481027172925989)
(7.15507155071551,0.478499483141385)
(7.29007290072901,0.47207504651515)
(7.42507425074251,0.461753863047285)
(7.56007560075601,0.44753593273779)
(7.69507695076951,0.429421255586665)
(7.83007830078301,0.407409831593909)
(7.96507965079651,0.381501660759523)
(8.10008100081001,0.351696743083507)
(8.23508235082351,0.31799507856586)
(8.37008370083701,0.280396667206583)
(8.50508505085051,0.2389070376235)
(8.64008640086401,0.197705434351329)
(8.77508775087751,0.160400577920789)
(8.91008910089101,0.126992468331878)
(9.04509045090451,0.0974811055845981)
(9.18009180091801,0.0718664896789483)
(9.31509315093151,0.0501486206149284)
(9.45009450094501,0.0323274983925392)
(9.58509585095851,0.0184031230117803)
(9.72009720097201,0.00837549447265167)
(9.85509855098551,0.00224461277515332)
(9.99009990099901,1.04779192852737e-05)
(10.1251012510125,0)
(10.260102601026,0)
(10.3951039510395,0)
(10.530105301053,0)
(10.6651066510665,0)
(10.80010800108,0)
(10.9351093510935,0)
(11.070110701107,0)
(11.2051120511205,0)
(11.340113401134,0)
(11.4751147511475,0)
(11.610116101161,0)
(11.7451174511745,0)
(11.880118801188,0)
(12.0151201512015,0)
(12.150121501215,0)
(12.2851228512285,0)
(12.420124201242,0)
(12.5551255512555,0)
(12.690126901269,0)
(12.8251282512825,0)
(12.960129601296,0)
(13.0951309513095,0)
(13.230132301323,0)
(13.3651336513365,0)

};
\path [draw=white, fill opacity=0] (axis cs:0,0.5)--(axis cs:13.5,0.5);

\path [draw=white, fill opacity=0] (axis cs:13.5,0)--(axis cs:13.5,0.5);

\path [draw=white, fill opacity=0] (axis cs:0,0)--(axis cs:13.5,0);

\path [draw=white, fill opacity=0] (axis cs:0,0)--(axis cs:0,0.5);

\node[coordinate,pin=above:{$t_{\mathrm{on}}, \frac{\epsilon_{ss}}{2}$}] at (axis cs:1.5,0.2405) {};

\node[coordinate,pin=above:{$t_{\mathrm{off}}, \frac{\epsilon_{ss}}{2}$}] at (axis cs:8.5,0.2405) {};

\node[coordinate,pin=below:{$\epsilon_{ss}$}] at (axis cs:5,0.4811) {};

\draw [dashed] (axis cs:7.0,0.0)--(axis cs:7.0,0.5);
\draw [dashed] (axis cs:0.0,0.0)--(axis cs:0.0,0.5);
\draw [dashed] (axis cs:3.0,0.0)--(axis cs:3.0,0.5);
\draw [dashed] (axis cs:10.0,0.0)--(axis cs:10.0,0.5);

\draw[stealth-stealth] (axis cs:0.0,0.4)--(axis cs:3.0,0.4);
\node[coordinate,pin=above:{$\sigma$}] at (axis cs:1.5,0.4) {};
\draw[stealth-stealth] (axis cs:7.0,0.4)--(axis cs:10.0,0.4);
\node[coordinate,pin=above:{$\sigma$}] at (axis cs:8.5,0.4) {};

\end{axis}

\end{tikzpicture}
\caption{Piecewise-quadratic input pulse, described by $t_{\mathrm{on}}$/$t_{\mathrm{off}}$ (times at which the measurement is turned on and off), $\sigma$ (the rise/decay time of the pulse), and $\epsilon_{\mathrm{ss}}$, the steady-state amplitude. For the remainder of the article, we set $\left( t_{\mathrm{on}}, \, t_{\mathrm{off}}, \, \sigma, \, \epsilon_{ss} \right) = \left( \nicefrac{1.5}{\chi}, \, \nicefrac{8.5}{\chi}, \, \nicefrac{3}{\chi}, \nicefrac{0.4811}{\sqrt{\chi}} \right)$. An additional time $\nicefrac{3.5}{\chi}$ is appended, to allow photons to exit the resonator.}
\label{fig:pulse}
\end{figure}
\begin{figure}[h!]
\centering
\input{Trace_Distance.tex}
\caption{Increase of trace distance between ideal post-measurement states as a result of pulse turn-off, in the three-qubit parity measurement described in \cite{SolgunDiV, TBDiV}. Pulse is shown in Fig. (\ref{fig:pulse}). System parameters are, in units of $\chi$: $\Delta_0 = 3$, $\Delta_1 = -3$, $\kappa_0 = \kappa_1 = 2$. No intrinsic sources of decoherence have been included. }
\label{fig:trace_distance}
\end{figure}
Since the trace distance is a measure of state distinguishability, this increase implies that the ability of this measurement to produce high-accuracy post-measurement states is higher than an otherwise-identical Markovian measurement would allow. 

\subsection{Stochastic}
The stochastic master equation governing homodyne measurement differs from the deterministic master equation only in the term $\meas{a_{\mathrm{out}}e^{-i\phi}}{\rho}dW$, given in Eq. (\ref{eq:m_c_rho}). Upon tracing out the resonator, the stochastic term in the master equation becomes:
\begin{flalign}
& \tr_{C}\left( \mathcal{M}[a_{\textrm{out}} e^{-i\phi}](\rho) \right) dW\nonumber \\
& = \tr_C \left( a_{\textrm{out}} e^{-i\phi}\rho + \rho a_{\textrm{out}}^{\dagger} e^{i\phi} - \left \langle a_{\textrm{out}} e^{-i\phi} + a_{\textrm{out}}^{\dagger} e^{i\phi} \right \rangle \rho \right)dW
\end{flalign}
Since the third term contains a full trace, we calculate it first:
\begin{flalign}
& \left \langle a_{\textrm{out}} e^{-i\phi} + a_{\textrm{out}}^{\dagger} e^{i\phi} \right \rangle \nonumber \\
& = \sum_{k}\sqrt{\kappa_k} \tr_{Q}\left( \tr_{C} \left( a_{\textrm{out}} e^{-i\phi}\rho + \rho a_{\textrm{out}}^{\dagger} e^{i\phi} \right) \right) \nonumber \\
& = \tr \left( \left( c_Q + c_Q \ct \right) \rho_Q \right) \label{eq:stoc-full-trace} \\
& \textrm{ where } c_Q \triangleq e^{-i\phi} \sum_{k}\sqrt{\kappa_k} \alpha_{k, i} \proj{i} \nonumber
\end{flalign}
This is also the expected value of the photocurrent in Eq. (\ref{eq:photocurrent}). We calculate the other two terms simultaneously, using the notation of Eq. (\ref{eq:rho_scal_func}) for concision:
\begin{flalign}
&\tr_C \left( e^{-i\phi} a_{\textrm{out}} \rho + e^{i\phi} \rho a_{\textrm{out}}^{\dagger} \right) \nonumber \\
&= \sum_k \sqrt{\kappa_k} \rho_Q \left(\tr\left( e^{-i\phi} a_k \ketbra{\alpha_{k,i}}{\alpha_{k,j}} + e^{i\phi} \ketbra{\alpha_{k,i}}{\alpha_{k,j}} a_k^{\dagger} \right)\right) \nonumber \\
&= \sum_k \sqrt{\kappa_k} \sum_{i,j} \left( \alpha_{k,i} e^{-i\phi} + \bar{\alpha}_{k,j} e^{i\phi} \right) \rho_{Q\,i,j} \ketbra{i}{j} \nonumber \\
&= c_Q\rho_Q + \rho_Q c_Q^{\dagger}
\end{flalign}
We see that $\tr_C \left( \mathcal{M}\left[a_{\textrm{out}}e^{-i\phi}\right](\rho)\right) = \mathcal{M}[c_Q](\rho_Q)$. Since $c_Q$ is diagonal, it commutes with the Hamiltonian, and is identical in the rotating frame, yielding the following stochastic master equation:
\begin{flalign}
\label{eq:cond-master-eqn}
d\rho_Q = & \dfrac{1}{2} \sum_{l \in Q} \gamma_{z,l} \diss{\sigma_{z,l}}{\rho_Q}dt
 -i \sum_{k\in C, l \in Q} \chi_{k,l} \tilde{P}_k \circ \com{\sigma_{z,l}}{\rho_Q}dt \nonumber \\
 &+ \sqrt{\eta}\meas{c_Q}{\rho_Q}dW.
\end{flalign}
In the following section, we numerically integrate the stochastic master equation, using the resulting state fidelity as a measure of performance.
\section{Simulation}
\label{sec:sims}
In order to assess the accuracy with which joint measurements can be made directly, we focus on the three-qubit parity measurement from \cite{SolgunDiV, TBDiV}. We set the resonator parameters according to Eq. (\ref{eq:deltas}) (with both $\kappa_0$ and $\kappa_1$ set to $2\chi$). 
\subsection{Methods}
We simulate the evolution of $\rho_Q$ over the interval $\left[t,t+dt\right]$ in two steps: we first determine the time-dependent amplitudes $\set{\alpha_{k,j}(t)}$ through numerical integration, using a $4^{\mathrm{th}}$/$5^{\mathrm{th}}$-order adaptive Runge-Kutta stepper, using the pulses detailed in Fig. (\ref{fig:pulse}), producing the state-dependent response shown in Fig. (\ref{fig:response}).
\begin{figure}[h!]
\centering
\input{Butterfly_Plot.tex}
\caption{Resonator responses to pulse in Fig. (\ref{fig:pulse}). Markers point in the direction of increasing time. The same output is seen in the steady state for bitstrings of identical parity, but the distinguishable transients reveal additional information about the Hamming weight.}
\label{fig:response}
\end{figure}
We then use these values of $\set{\alpha_{k,j}(t)}$ to formulate the time-dependent reduced master equation (Eq. (\ref{eq:cond-master-eqn})), and use an order-1.5 stochastic Runge-Kutta method \cite{NumericalSolutionSDEs} to integrate it. We repeat this for $10^5$ uniformly-spaced timesteps on the interval $\left \lbrack 0, \tau \right \rbrack$, where $\tau$ is the total measurement time (taken to be $\nicefrac{13.5}{\chi}$ throughout). 

In order to verify the correctness of the above simulation, we calculate the minimum/maximum eigenvalues, traces, deviations from hermiticity $\left( \norm{\infty}{\rho - \rho^{\dagger}} \right)$ and purities $\left( \tr \left(\rho^2 \right)\right)$ for a typical trajectory. In order for these quantities to be meaningful when plotted for a single trajectory, the algorithm used has to be strong (in the terminology of stochastic differential equations \cite{NumericalSolutionSDEs}). Deviations from hermiticity on the order $10^{-15}$ are typical, as are deviations from unit trace on the order $10^{-13}$, likely caused by numerical rounding error. For the simulations discussed in this paper, the remaining checks are satisfied to within $\sim 10^{-49}$.
\subsection{Figure of Merit}
To calculate the performance of the continuous measurement, we evaluate its ability to produce one of two definite-parity entangled states, projected from an initial state $\ket{+++}$, and to produce a measurement record which correctly identifies the parity of the final state. To this end, we calculate the quantum state fidelity:
\begin{equation}
F_{\pm} = \sqrt{\elem{\psi_{\pm}}{\rho_Q}{\psi_{\pm}}},
\end{equation}
where $\ket{\psi_{\pm}}$ is one of the parity eigenstates from Eq. (\ref{eq:post-measurement-states}). This fidelity is calculated for a post-selected ensemble determined by the signal $s(\tau)$, which is a weighted integral of the measurement record:
\begin{equation}
s(\tau) = \int_{0}^{\tau} f(t) j(t) dt.
\end{equation}
If this signal is positive (negative), we infer that the parity of the post-measurement state is even (odd). 

The choice of filter function $f(t)$ has a significant impact on the performance of the measurement. Though it has no effect on the post-measurement states themselves, it can increase or decrease the probability of an incorrect assignment of states to ensembles, affecting the fidelity indirectly. We compare (in Fig. (\ref{fig:pc_hist})) the distributions of $s(\tau)$ corresponding to the uniform filter $\left( f(t) = 1 \right)$ and a matched filter (a filter function proportional to the measurement record, see \cite{DetectionEstimationModulation, OptimalFiltering}), shown in Fig. (\ref{fig:matched_filter}).
\begin{figure}[h!]
\centering
\input{PC_Histogram.tex}
\caption{Histogram of $10010$ integrated photocurrents corresponding to the input pulse in Fig. (\ref{fig:pulse}). }
\label{fig:pc_hist}
\end{figure}
\begin{figure}[h!]
\centering
\input{MatchedFilter.tex}
\caption{The approximate matched filter given by Eq. (\ref{eq:filter_function}), for use in obtaining a measurement signal $s(\tau)$ given a photocurrent time trace $j(t)$. The filter function can be thought of as assigning an importance to the photocurrent emitted at time $t$; this filter function assigns high importance during the steady-state phase of the measurement, and low importance at other times.}
\label{fig:matched_filter}
\end{figure}

This matched filter is derived from a simplified model of the dynamics, in which we assume that the state immediately collapses to a uniform mixture of the computational basis states from one of the parity eigenspaces. The filter is then the expected value of the nominal photocurrent in the even-parity subspace, normalized to have a mean value of 1:
\begin{equation}
\label{eq:filter_function}
f(t) = \frac{j_+(t)}{\int_{0}^{\tau} j_+(t')dt'},
\end{equation}
where $j_+(t) = \tr \left( \left(c_Q + c_Q \ct\right) \Pi_{+} \right)$ (see Eq. (\ref{eq:stoc-full-trace})). Here, we have defined $\Pi_{\pm}$ to be projectors onto the even/odd parity subspaces of $B$, and we have taken advantage of the fact that $\tr \left( \left(c_Q + c_Q \ct\right) \Pi_{-} \right) = -\tr \left( \left(c_Q + c_Q \ct\right) \Pi_{+} \right)$. In the following section, we use the post-selected state fidelity to examine the performance of a nominal measurement.
\subsection{Results \& Discussion}
We simulate the action of the measurement in the presence of decoherence from \chg{the measurement itself and qubit dephasing noise}. Selecting $\chi$ as the natural scale for frequency, we take $\gamma_{z}$ to be $\nicefrac{\chi}{300}$, as in \cite{TBDiV}. If $\chi \sim 1 \times 2\pi $ MHz, this would correspond to a dephasing time of $\sim 50 \mu$s.

A histogram of integrated photocurrents is given in Fig. (\ref{fig:pc_hist}). The separation between the two Gaussian peaks therein is visibly greater when using the approximate matched filter from Eq. (\ref{eq:filter_function}), indicating that the probability of incorrect assignment is decreased. The resulting state fidelities are given in Fig. (\ref{fig:fid_hist}).
\begin{figure}[h!]
\centering
\input{Fidelity_Histogram.tex}
\caption{Histogram of $5031$ post-selected state fidelities corresponding to events in which odd parity is detected $\left( s(\tau)<0 \right)$. State fidelity of the average post-selected state (given by the root mean square of the distribution above) is $\sim 94\%$.}
\label{fig:fid_hist}
\end{figure}
To understand the features of the fidelity distribution, we plot (in Fig. (\ref{fig:det_fid})) the expected fidelity for a simplified model of measurement in which the state immediately collapses to one of the parity eigenstates, and is then subject to both intrinsic and measurement-induced decoherence.
\begin{figure}[h!]
\centering
\input{Four_State_Fidelities.tex}
\caption{State fidelity for a simplified model of measurement, in which the state immediately collapses to the ideal post-measurement state, and is then acted upon by intrinsic decoherence and/or the coupling Lindbladian. The decay in the fidelity of the state $\frac{1}{\sqrt{3}}\left( \ket{100} + \ket{010} + \ket{001} \right)$ without the coupling Lindbladian (solid black line) coincides exactly with the case in which the coupling Lindbladian is present (black circles). The initial state $\frac{1}{2}\left( \ket{100} + \ket{100} + \ket{100}   + \ket{111}\right)$ produces a markedly different effect, its fidelity decaying much more quickly when the coupling lindbladian is present (solid blue line) than when it is absent (dashed black line).}
\label{fig:det_fid}
\end{figure}

In order to compare this average state fidelity with known performance thresholds for error-correcting architectures \cite{FaultTolerantArchitectures}, we calculate the post-measurement state fidelity obtained by performing a circuit-based measurement \cite{DKLP}. This measurement uses the circuit in Fig. (\ref{fig:measurement_circuit}).
\begin{figure}[h!]
\centering
\mbox{
\Qcircuit @C=1em @R=1em {
                 & \ctrl{3} & \qw      & \qw      & \qw    \\
                 & \qw      & \ctrl{2} & \qw      & \qw    \\
                 & \qw      & \qw      & \ctrl{1} & \qw    \\
\lstick{\ket{0}} & \targ    & \targ    & \targ    & \meter 
}
}
\caption{A measurement circuit for the operator $ZZZ$, similar to the four-qubit circuit used in \cite{DKLP}.}
\label{fig:measurement_circuit}
\end{figure}

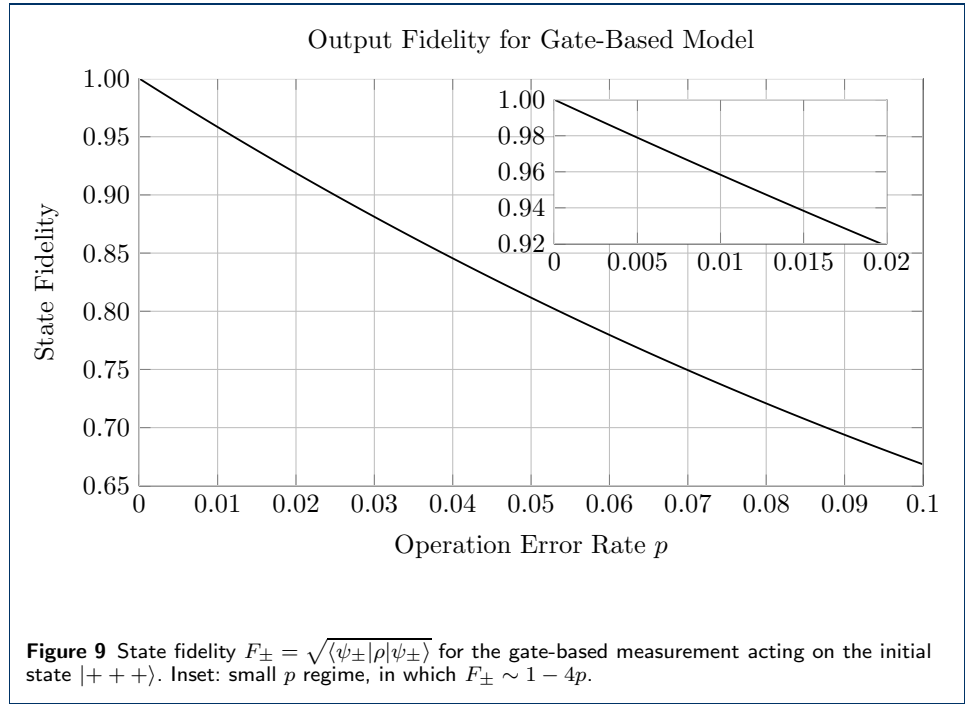
\begin{figure}[h!]
\centering
%
%
%
%
\begin{tikzpicture}

\definecolor{color0}{rgb}{0.917647058823529,0.917647058823529,0.949019607843137}

\begin{axis}[
title={Output Fidelity for Gate-Based Model},
xlabel={Operation Error Rate $p$},
ylabel={State Fidelity},
xmin=0, xmax=0.1,
ymin=0.65, ymax=1,
width=\figurewidth,
height=\figureheight,
xtick={0,0.01,0.02,0.03,0.04,0.05,0.06,0.07,0.08,0.09,0.1},
xticklabels={0,0.01,0.02,0.03,0.04,0.05,0.06,0.07,0.08,0.09,0.1},
ytick={0.65,0.7,0.75,0.8,0.85,0.9,0.95,1},
yticklabels={0.65,0.70,0.75,0.80,0.85,0.90,0.95,1.00},
xmajorgrids,
ymajorgrids,
]
\addplot [line width=0.7000000000000001pt, black]
coordinates {
(0,1)
(0.00101010101010101,0.995700967493982)
(0.00202020202020202,0.991423355178075)
(0.00303030303030303,0.98716709801483)
(0.00404040404040404,0.982932131084236)
(0.00505050505050505,0.978718389582822)
(0.00606060606060606,0.974525808822745)
(0.00707070707070707,0.970354324230877)
(0.00808080808080808,0.966203871347883)
(0.00909090909090909,0.962074385827289)
(0.0101010101010101,0.957965803434552)
(0.0111111111111111,0.953878060046119)
(0.0121212121212121,0.949811091648474)
(0.0131313131313131,0.945764834337191)
(0.0141414141414141,0.941739224315973)
(0.0151515151515152,0.937734197895683)
(0.0161616161616162,0.933749691493376)
(0.0171717171717172,0.929785641631323)
(0.0181818181818182,0.925841984936024)
(0.0191919191919192,0.921918658137222)
(0.0202020202020202,0.918015598066906)
(0.0212121212121212,0.914132741658316)
(0.0222222222222222,0.910270025944933)
(0.0232323232323232,0.906427388059471)
(0.0242424242424242,0.902604765232862)
(0.0252525252525253,0.898802094793234)
(0.0262626262626263,0.895019314164884)
(0.0272727272727273,0.891256360867252)
(0.0282828282828283,0.887513172513882)
(0.0292929292929293,0.883789686811386)
(0.0303030303030303,0.880085841558397)
(0.0313131313131313,0.876401574644521)
(0.0323232323232323,0.872736824049287)
(0.0333333333333333,0.869091527841091)
(0.0343434343434343,0.865465624176133)
(0.0353535353535354,0.861859051297355)
(0.0363636363636364,0.858271747533374)
(0.0373737373737374,0.854703651297414)
(0.0383838383838384,0.851154701086228)
(0.0393939393939394,0.847624835479025)
(0.0404040404040404,0.844113993136388)
(0.0414141414141414,0.840622112799197)
(0.0424242424242424,0.837149133287539)
(0.0434343434343434,0.833694993499628)
(0.0444444444444444,0.830259632410711)
(0.0454545454545455,0.826842989071982)
(0.0464646464646465,0.823445002609491)
(0.0474747474747475,0.820065612223046)
(0.0484848484848485,0.816704757185127)
(0.0494949494949495,0.813362376839784)
(0.0505050505050505,0.810038410601547)
(0.0515151515151515,0.806732797954329)
(0.0525252525252525,0.803445478450328)
(0.0535353535353535,0.800176391708934)
(0.0545454545454545,0.796925477415632)
(0.0555555555555556,0.793692675320911)
(0.0565656565656566,0.790477925239164)
(0.0575757575757576,0.787281167047604)
(0.0585858585858586,0.784102340685163)
(0.0595959595959596,0.780941386151412)
(0.0606060606060606,0.777798243505467)
(0.0616161616161616,0.77467285286491)
(0.0626262626262626,0.771565154404701)
(0.0636363636363636,0.768475088356102)
(0.0646464646464646,0.765402595005599)
(0.0656565656565657,0.762347614693833)
(0.0666666666666667,0.759310087814527)
(0.0676767676767677,0.756289954813422)
(0.0686868686868687,0.75328715618722)
(0.0696969696969697,0.750301632482524)
(0.0707070707070707,0.747333324294795)
(0.0717171717171717,0.744382172267297)
(0.0727272727272727,0.741448117090067)
(0.0737373737373737,0.73853109949888)
(0.0747474747474747,0.735631060274221)
(0.0757575757575758,0.732747940240264)
(0.0767676767676768,0.729881680263869)
(0.0777777777777778,0.727032221253564)
(0.0787878787878788,0.724199504158561)
(0.0797979797979798,0.721383469967759)
(0.0808080808080808,0.718584059708771)
(0.0818181818181818,0.715801214446947)
(0.0828282828282828,0.713034875284419)
(0.0838383838383838,0.710284983359146)
(0.0848484848484849,0.707551479843973)
(0.0858585858585859,0.704834305945703)
(0.0868686868686869,0.702133402904176)
(0.0878787878787879,0.699448711991361)
(0.0888888888888889,0.696780174510458)
(0.0898989898989899,0.694127731795016)
(0.0909090909090909,0.691491325208062)
(0.0919191919191919,0.688870896141239)
(0.0929292929292929,0.686266386013961)
(0.0939393939393939,0.683677736272585)
(0.094949494949495,0.681104888389589)
(0.095959595959596,0.678547783862771)
(0.096969696969697,0.67600636421446)
(0.097979797979798,0.673480570990743)
(0.098989898989899,0.670970345760709)
(0.1,0.668475630115706)

};
\path [draw=white, fill opacity=0] (axis cs:0,1)--(axis cs:0.1,1);

\path [draw=white, fill opacity=0] (axis cs:0.1,0.65)--(axis cs:0.1,1);

\path [draw=white, fill opacity=0] (axis cs:0,0.65)--(axis cs:0.1,0.65);

\path [draw=white, fill opacity=0] (axis cs:0,0.65)--(axis cs:0,1);

\end{axis}

\begin{axis}[
axis background/.style={fill=white},
xmin=0, xmax=0.02,
ymin=0.92, ymax=1,
width=0.5\figurewidth,
height=0.5\figureheight,
xtick={0,0.005,0.01,0.015,0.02},
xticklabels={0,0.005,0.01,0.015,0.02},
scaled x ticks=false,
ytick={0.92,0.94,0.96,0.98,1},
yticklabels={0.92,0.94,0.96,0.98,1.00},
xmajorgrids,
ymajorgrids,
xshift=0.46\figurewidth,
yshift=0.46\figureheight,
x tick label style={/pgf/number format/fixed,}
]
\addplot [line width=0.7000000000000001pt, black]
coordinates {
(0,1)
(0.000202020202020202,0.999138476757821)
(0.000404040404040404,0.99827781240747)
(0.000606060606060606,0.997418006427512)
(0.000808080808080808,0.996559058296701)
(0.00101010101010101,0.995700967493982)
(0.00121212121212121,0.994843733498488)
(0.00141414141414141,0.993987355789544)
(0.00161616161616162,0.993131833846661)
(0.00181818181818182,0.992277167149542)
(0.00202020202020202,0.991423355178075)
(0.00222222222222222,0.990570397412339)
(0.00242424242424242,0.989718293332599)
(0.00262626262626263,0.988867042419309)
(0.00282828282828283,0.98801664415311)
(0.00303030303030303,0.98716709801483)
(0.00323232323232323,0.986318403485482)
(0.00343434343434343,0.985470560046269)
(0.00363636363636364,0.984623567178577)
(0.00383838383838384,0.98377742436398)
(0.00404040404040404,0.982932131084236)
(0.00424242424242424,0.982087686821289)
(0.00444444444444444,0.981244091057268)
(0.00464646464646465,0.980401343274487)
(0.00484848484848485,0.979559442955444)
(0.00505050505050505,0.978718389582822)
(0.00525252525252525,0.977878182639485)
(0.00545454545454545,0.977038821608485)
(0.00565656565656566,0.976200305973054)
(0.00585858585858586,0.975362635216608)
(0.00606060606060606,0.974525808822745)
(0.00626262626262626,0.973689826275246)
(0.00646464646464646,0.972854687058075)
(0.00666666666666667,0.972020390655375)
(0.00686868686868687,0.971186936551473)
(0.00707070707070707,0.970354324230877)
(0.00727272727272727,0.969522553178275)
(0.00747474747474747,0.968691622878535)
(0.00767676767676768,0.967861532816707)
(0.00787878787878788,0.96703228247802)
(0.00808080808080808,0.966203871347883)
(0.00828282828282828,0.965376298911883)
(0.00848484848484848,0.964549564655788)
(0.00868686868686869,0.963723668065545)
(0.00888888888888889,0.962898608627278)
(0.00909090909090909,0.962074385827289)
(0.00929292929292929,0.961250999152059)
(0.0094949494949495,0.960428448088246)
(0.0096969696969697,0.959606732122687)
(0.0098989898989899,0.958785850742392)
(0.0101010101010101,0.957965803434552)
(0.0103030303030303,0.957146589686533)
(0.0105050505050505,0.956328208985874)
(0.0107070707070707,0.955510660820295)
(0.0109090909090909,0.954693944677687)
(0.0111111111111111,0.953878060046119)
(0.0113131313131313,0.953063006413833)
(0.0115151515151515,0.952248783269248)
(0.0117171717171717,0.951435390100953)
(0.0119191919191919,0.950622826397716)
(0.0121212121212121,0.949811091648474)
(0.0123232323232323,0.949000185342341)
(0.0125252525252525,0.948190106968602)
(0.0127272727272727,0.947380856016715)
(0.0129292929292929,0.946572431976311)
(0.0131313131313131,0.945764834337191)
(0.0133333333333333,0.944958062589332)
(0.0135353535353535,0.944152116222878)
(0.0137373737373737,0.943346994728147)
(0.0139393939393939,0.942542697595625)
(0.0141414141414141,0.941739224315973)
(0.0143434343434343,0.940936574380018)
(0.0145454545454545,0.940134747278759)
(0.0147474747474747,0.939333742503363)
(0.0149494949494949,0.938533559545169)
(0.0151515151515152,0.937734197895683)
(0.0153535353535354,0.936935657046579)
(0.0155555555555556,0.936137936489701)
(0.0157575757575758,0.935341035717061)
(0.015959595959596,0.934544954220837)
(0.0161616161616162,0.933749691493376)
(0.0163636363636364,0.932955247027192)
(0.0165656565656566,0.932161620314965)
(0.0167676767676768,0.931368810849541)
(0.016969696969697,0.930576818123935)
(0.0171717171717172,0.929785641631323)
(0.0173737373737374,0.928995280865052)
(0.0175757575757576,0.92820573531863)
(0.0177777777777778,0.927417004485731)
(0.017979797979798,0.926629087860195)
(0.0181818181818182,0.925841984936024)
(0.0183838383838384,0.925055695207387)
(0.0185858585858586,0.924270218168612)
(0.0187878787878788,0.923485553314195)
(0.018989898989899,0.922701700138791)
(0.0191919191919192,0.921918658137222)
(0.0193939393939394,0.921136426804468)
(0.0195959595959596,0.920355005635673)
(0.0197979797979798,0.919574394126143)
(0.02,0.918794591771345)

};
\path [draw=white, fill opacity=0] (axis cs:0,1)--(axis cs:0.02,1);

\path [draw=white, fill opacity=0] (axis cs:0.02,0.92)--(axis cs:0.02,1);

\path [draw=white, fill opacity=0] (axis cs:0,0.92)--(axis cs:0.02,0.92);

\path [draw=white, fill opacity=0] (axis cs:0,0.92)--(axis cs:0,1);

\end{axis}

\end{tikzpicture}
\caption{State fidelity $F_{\pm} = \sqrt{ \bra{\psi_{\pm}} \rho \ket{\psi_{\pm}} } $ for the gate-based measurement acting on the initial state $\ket{+++}$. Inset: small $p$ regime, in which $F_{\pm} \sim 1 - 4p$.}
\label{fig:analytic_fidelity}
\end{figure}

We assign to each operation in the circuit (state preparation, memory, \textsc{CNOT} and single-qubit measurement) a failure probability $p$. Failure corresponds to:
\begin{itemize}
\item unintentional preparation of the $\ket{1}$ state on the ancilla qubit,
\item insertion of a random one-qubit Pauli operator ($X$, $Y$, or $Z$) after a memory operation with probability $\nicefrac{1}{3}$,
\item insertion of a random two-qubit Pauli operator ($IX$, $IY$, \ldots, $ZZ$) after a \textsc{CNOT} operation with probability $\nicefrac{1}{15}$, and
\item an incorrect ancilla measurement outcome  
\end{itemize}
for the four basic operations, respectively. The resulting state fidelity can be calculated exactly:
\begin{flalign}
F_{\pm}(p) =  \Bigg[1 &- \frac{128}{15} p + \frac{8834}{225} p^{2} - \frac{74884}{675} p^{3} + \frac{2130272}{10125} p^{4} - \frac{8409088}{30375}  p^{5}  \nonumber \\ 
& + \frac{23153152}{91125} p^{6} - \frac{43695104}{273375} p^{7} + \frac{53886976}{820125} p^{8} - \frac{39059456}{2460375} p^{9} \nonumber \\ 
&  + \frac{4194304}{2460375} p^{10} \Bigg]^{\nicefrac{1}{2}}
\end{flalign}
Plotting this fidelity in Fig. (\ref{fig:analytic_fidelity}), we see that a state fidelity of $94\%$ corresponds to an error rate of $1.4-1.5\%$, and that an output state fidelity of $96-98\%$ would match the performance of the gate-based model close to the error-correction threshold. This indicates that, given the parameters for existing state-of-the-art transmon/cavity systems, multi-qubit measurements of quality near that required for fault-tolerant error correction could be performed, but that further improvements in control design and/or hardware parameters would be needed for threshold error rates to be attained.
\section{Conclusions and Future Work}
\label{sec:conc}
The formalism presented in this article facilitates the design of a class of quantum measurement devices, with applications in fault-tolerant quantum computing architectures and remote entanglement preparation. Using a pulse with few free parameters, and an approximate matched filter, it is possible to limit transient-induced decoherence, achieving a high state fidelity, comparable to error models studied in the fault tolerance literature operating slightly above the fault-tolerance threshold. Further advances in the design of quantum hardware, such as decreases in the dephasing rate, will permit higher-fidelity implementations of this protocol. In addition, there are several purely theoretical avenues to be explored, which can inform the feasibility of this idea. 

In future work, we will attempt to eliminate or minimize the decoherent portion of the coupling Lindbladian, by selecting a control pulse optimally. Such a control pulse can have a quadrature-phase component, unlike the pulse used in this manuscript. Decreasing this decoherence will permit stronger driving, which in turn enables shorter measurement times, reducing the effective strength of intrinsic decoherence. We will also attempt to minimize or correct unwanted rotations, which occur as a result of the ac-Stark shift and the non-zero imaginary part of $c_Q$. It is known that non-linear filtering of the output photocurrent and multi-qubit gates can be used for this purpose \cite{UndoingDephasing}, but the performance of a scheme involving efficient filters and single-qubit gates has yet to be examined.

\section{Remarks}
Numerical simulations were performed using the libraries \href{http://github.com/bcriger/homodyne_sim/}{\texttt{homodyne\_sim}} and \href{http://github.com/bcriger/sde_solve/}{\texttt{sde\_solve}}, available on Github.
\begin{appendices}
\section{Full Dispersive Hamiltonian and Purcell Term}
\label{appa}

Written in the laboratory frame, the multi-qubit, multi-mode Jaynes-Cummings Hamiltonian is
\begin{equation}
 H_{JC}= \sum_{l \in Q} \frac{\hbar \Omega_{l}}{2} \sigma_{z,l} + \sum_{k \in C} \hbar \omega_k a_{k}^{\dagger} a_k+\sum_{l \in Q}  \sum_{k \in C} \hbar g_{k,l}(a_k^{\dagger} \sigma_{l}^- + \sigma_l^{+} a_k) 
 \end{equation}
This expression introduces the mode- and qubit-dependent couplings $g_{k,l}$.  Recall that a Rotating Wave Approximation (RWA) has already been made in the form of the mode-qubit coupling  \footnote{A formal derivation of the interaction Hamiltonian between two cavity modes can be found in \cite{steckNotes}.}. 

Applying the standard canonical-transformation (or Schrieffer-Wolff) analysis to this Hamiltonian in the dispersive regime, i.e., $|g_{k,l}| \ll |\Omega_l-\omega_k |$ for all $k$ and $l$, one obtains the effective Hamiltonian, to second order in $\nicefrac{g_{k,l}}{\Delta_{k,l}}$ \cite{PhysRevA.69.062320}:
 \begin{multline}
 H_{JC,d} = \sum_{l \in Q} \hbar \biggl (\frac{\Omega_l}{2} + \sum_{k \in C} \frac{g_{k,l}^2}{\Delta_{k,l}} \biggr) \sigma_{z,l}  + \sum_{k \in C} \hbar \biggl (\omega_k + \sum_{l \in Q} \frac{g_{k,l}^2}{\Delta_{k,l}} \sigma_{z,l} \biggr) a_{k}^{\dagger} a_k \\
+\frac{\hbar}{4} \sum_{\substack{l,l \sp{\prime} \in Q \\ l \neq l  \sp{\prime}}} \biggl ( \sum_{k \in C} \frac{g_{k,l} g_{k, l \sp{\prime}}(\Delta_{k,l} + \Delta_{k, l \sp{\prime}})}{\Delta_{k,l} \Delta_{k, l \sp{\prime}}} \biggr)(\sigma_{l}^+ \sigma_{l  \sp{\prime}}^-+\sigma_{l  \sp{\prime}}^+ \sigma_l^-) \\
+ \frac{\hbar}{4} \sum_{\substack{k, k  \sp{\prime} \in C \\ k \neq k  \sp{\prime}}} \biggl ( \sum_{l \in Q} \frac{g_{k, l} g_{k  \sp{\prime}, l} (\Delta_{k,l}+\Delta_{k  \sp{\prime}, l})}{\Delta_{k,l} \Delta_{k, l  \sp{\prime}}} \sigma_{z,l} \biggr) (a_k^{\dagger} a_{k  \sp{\prime}} + a_{k  \sp{\prime}}^{\dagger} a_k ).\label{SWtrans}
 \end{multline}
 Here $\Delta_{k,l}\triangleq  \Omega_l-\omega_k$.  We note the following differences between Eq. (\ref{SWtrans}) and our model Hamiltonian Eq. (\ref{dispeq}) in the text:
 \begin{itemize}
 \item In the first (Lamb shift) and second (dispersive shift) terms, we identify the parameter $\chi_{k,l}$ with $g_{k,l}^2/\Delta_{k,l}$; this is perfectly conventional.
 \item The third term is a new qubit-qubit coupling term \cite{InPreparation}.  As we discuss in the text, the parity measurement is compatible with all the qubits being far apart (many $\chi$) in frequency; thus it is normal to invoke an RWA to neglect this term.
 \item The final term is a new conditional mode-mode coupling term.
 \end{itemize}
 This last term requires some further discussion.  Note that since this term commutes with $\sigma_{z,l}$, it is compatible with a nondemolition measurement of the qubit states; more than that, its form is compatible with the view that the state of the qubits determines the value of the ``dielectric constant'' of the linear resonator system, whose frequency can be probed by the phase shift of scattered coherent radiation.  From this point of view, this mode-mode coupling does not alter the fundamental strategy, or the basic formalism, of the parity-measurement setup.  
 
 However, this term does produce some new complications.  The response is now not only determined by the $\chi_{k,l}$ parameters, but depends separately on the $g$ and $\Delta$ parameters.  The response is not automatically determined by only the Hamming weight of the state of the qubits; but the difference in the response of states with the same Hamming weight can be made weak if the overall detuning is large compared with the differences of qubit frequencies.  If the $\kappa$ parameters of the two modes are made significantly different, then, according to Eqs. (\ref{eq:deltas}), the bare resonator frequencies can be made significantly different; this in turn will make the influence of the mode-mode coupling term very small, in accordance with another RWA.  
 
 We will report in a forthcoming publication \cite{InPreparation} on the re-optimization of the parity measurement with all the effects of this Hamiltonian Eq. (\ref{SWtrans}) taken into account.  It is especially worthwhile to do this, in light of the fact that a similar analysis of the dissipative terms shows that we can design a structure which both does a good parity measurement and has built-in protection against Purcell decay.  The relevant decay term can be written \cite{InPreparation}:
 \begin{equation}
 \mathcal{D}\left[\sum_{l \in Q} \biggl(\sum_{k \in C} \frac{g_{k,l}}{\Delta_{k,l}} \sqrt{\kappa_k}\biggr) \sigma_l^{-}\right](\rho).
\end{equation}
The inner sum over the cavity modes gives the opportunity for cancellation of the coefficient of each $\sigma^-_l$ term, since the $g$ factors can be set to have either sign.  Our preliminary work indicates that this indeed can be done, in a way that is consistent with satisfying the conditions for the parity measurement.  We will provide the details of this analysis in an upcoming publication.
 
\section{Pointer State Solution}
\label{appb}

Here we sketch the proof that the pointer states in Eq. (\ref{eq:state_ansatz}) constitute a family of solutions of the multi-qubit multi-mode master equation (Eq. \ref{eq:me}), when no qubit damping is present $\gamma_{-,l}=\lambda_{k,l}=0$.  First, one establishes by direct substitution that the ``diagonal states''
\begin{equation}
\rho =  \ketbra{i}{i} \otimes \bigotimes_{k \in C} \ketbra{\alpha_{k,i}(t)}{\alpha_{k,i}(t)},
\end{equation}
are a solution to Eq. (\ref{eq:me}), where the functions $\alpha_{k,i}(t)$ are the solutions to Eq. (\ref{eq:cavity_eom}).  This is established by observing that the entire right-hand side of the equation commutes with $\sigma_{z,l}$.  The form of the remaining terms (involving the off-diagonal qubit operator $\ketbra{i}{j}$) does {\em not} follow from any simple linearity argument.  However, we can seek an extended solution having these new terms,
\begin{equation}
\rho =  \sum_i \rho_{ii}\ketbra{i}{i} \otimes \bigotimes_{k \in C} \ketbra{\alpha_{k,i}(t)}{\alpha_{k,i}(t)}+\sum_{i,j,i\neq j}\ketbra{i}{j} \otimes\hat O_{i,j}(t).
\end{equation}
Here $\hat O_{i,j}(t)$ is an undetermined operator on the Hilbert space of the resonators, which is a function of the register states $\ket{i}$ and $\ket{j}$, as well as being an undertermined function of time.  $\hat O$ is fixed by positivity; it can be shown (details in \cite{InPreparation}) that $\rho\ge 0$ forces $\hat O_{i,j}(t)$ to be of the form
\begin{equation}
\hat O_{i,j}(t)=\rho_{ij}(t)\bigotimes_{k \in C} \ketbra{\alpha_{k,i}(t)}{\alpha_{k,j}(t)}.
\end{equation}
This confirms Eq. (\ref{eq:state_ansatz}), with the added information that the diagonal coefficients are independent of time, which is a property of the equation of motion that we subsequently derive for $\rho_Q$ (Eq. (\ref{eq:unrot-frame-eqn})).
\end{appendices}

\begin{backmatter}

\section*{Competing interests}
  The authors declare that they have no competing interests.

\section*{Author's contributions}
    Derivations and numerical simulations were contributed by BC, non-Markovianity of the coupling Lindbladian was proved by DD. AC contributed the detailed analysis of the multi-qubit multi-mode Jaynes-Cummings hamiltonian. BC and DD contributed to the writing of this paper.

\section*{Acknowledgements}
The authors thank Shabir Barzanjeh, Felix Motzoi, Lars Tornberg, and Anna Vershynina for helpful discussions. The authors acknowledge financial support from ScaleQIT.

\bibliographystyle{vancouver} 
\bibliography{derivation_corrections} 

\begin{thebibliography}{10}

\bibitem{SingleQubitGates}
{Chow} JM, {Dicarlo} L, {Gambetta} JM, {Motzoi} F, {Frunzio} L, {Girvin} SM,
  et~al.
\newblock {Optimized driving of superconducting artificial atoms for improved
  single-qubit gates}.
\newblock Phys Rev A. 2010 Oct;82(4):040305.

\bibitem{TwoQubitGates}
{C{\'o}rcoles} AD, {Gambetta} JM, {Chow} JM, {Smolin} JA, {Ware} M, {Strand} J,
  et~al.
\newblock {Process verification of two-qubit quantum gates by randomized
  benchmarking}.
\newblock Phys Rev A. 2013 Mar;87(3):030301.

\bibitem{SingleShotMeasurement}
{Mallet} F, {Ong} FR, {Palacios-Laloy} A, {Nguyen} F, {Bertet} P, {Vion} D,
  et~al.
\newblock {Single-shot qubit readout in circuit quantum electrodynamics}.
\newblock Nature Physics. 2009 Nov;5:791--795.

\bibitem{NielsenChuang}
Nielsen MA, Chuang IL.
\newblock Quantum Computation and Quantum Information: 10th Anniversary
  Edition.
\newblock Cambridge University Press; 2010.
\newblock Available from: \url{https://books.google.de/books?id=-s4DEy7o-a0C}.

\bibitem{ToricCode}
{Kitaev} AY.
\newblock {Fault-tolerant quantum computation by anyons}.
\newblock Annals of Physics. 2003 Jan;303:2--30.

\bibitem{BaconShorCodes}
{Bacon} D.
\newblock {Operator quantum error-correcting subsystems for self-correcting
  quantum memories}.
\newblock Phys Rev A. 2006 Jan;73(1):012340.

\bibitem{ColourCodes}
{Bombin} H, {Martin-Delgado} MA.
\newblock {Topological Quantum Distillation}.
\newblock Physical Review Letters. 2006 Nov;97(18):180501.

\bibitem{SparseGraphCodes}
MacKay DJC, Mitchison G, McFadden PL.
\newblock Sparse-graph codes for quantum error correction.
\newblock Information Theory, IEEE Transactions on. 2004 Oct;50(10):2315--2330.

\bibitem{ShorEC}
Shor PW.
\newblock Fault-tolerant quantum computation.
\newblock In: Foundations of Computer Science, 1996. Proceedings., 37th Annual
  Symposium on; 1996. p. 56--65.

\bibitem{SteaneEC}
{Steane} AM.
\newblock {Active Stabilization, Quantum Computation, and Quantum State
  Synthesis}.
\newblock Physical Review Letters. 1997 Mar;78:2252--2255.

\bibitem{KnillEC}
{Knill} E.
\newblock {Quantum computing with realistically noisy devices}.
\newblock Nature. 2005 Mar;434:39--44.

\bibitem{DKLP}
{Dennis} E, {Kitaev} A, {Landahl} A, {Preskill} J.
\newblock {Topological quantum memory}.
\newblock Journal of Mathematical Physics. 2002 Sep;43:4452--4505.

\bibitem{SingleQubitHomodyneMeasurement}
{Gambetta} J, {Blais} A, {Boissonneault} M, {Houck} AA, {Schuster} DI, {Girvin}
  SM.
\newblock {Quantum trajectory approach to circuit QED: Quantum jumps and the
  Zeno effect}.
\newblock Phys Rev A. 2008 Jan;77(1):012112.

\bibitem{TunableJointMeasurement}
{Lalumi{\`e}re} K, {Gambetta} JM, {Blais} A.
\newblock {Tunable joint measurements in the dispersive regime of cavity QED}.
\newblock Phys Rev A. 2010 Apr;81(4):040301.

\bibitem{UndoingDephasing}
{Frisk Kockum} A, {Tornberg} L, {Johansson} G.
\newblock {Undoing measurement-induced dephasing in circuit QED}.
\newblock Phys Rev A. 2012 May;85(5):052318.

\bibitem{RemoteEntanglement}
Motzoi F, Whaley KB, Sarovar M.
\newblock Continuous joint measurement and entanglement of qubits in remote
  cavities.
\newblock Phys Rev A. 2015 Sep;92:032308.
\newblock Available from:
  \url{http://link.aps.org/doi/10.1103/PhysRevA.92.032308}.

\bibitem{SolgunDiV}
{DiVincenzo} DP, {Solgun} F.
\newblock {Multi-qubit parity measurement in circuit quantum electrodynamics}.
\newblock New Journal of Physics. 2013 Jul;15(7):075001.

\bibitem{TBDiV}
{Tornberg} L, {Barzanjeh} S, {DiVincenzo} DP.
\newblock {Stochastic-master-equation analysis of optimized three-qubit
  nondemolition parity measurements}.
\newblock Phys Rev A. 2014 Mar;89(3):032314.

\bibitem{ExploringTheQuantum}
Haroche S, Raimond JM.
\newblock Exploring the Quantum: Atoms, Cavities, and Photons.
\newblock Oxford Graduate Texts. OUP Oxford; 2013.
\newblock Available from: \url{https://books.google.de/books?id=YHC1kQEACAAJ}.

\bibitem{PhysRevA.69.062320}
Blais A, Huang RS, Wallraff A, Girvin SM, Schoelkopf RJ.
\newblock Cavity quantum electrodynamics for superconducting electrical
  circuits: An architecture for quantum computation.
\newblock Phys Rev A. 2004 Jun;69:062320.
\newblock Available from:
  \url{http://link.aps.org/doi/10.1103/PhysRevA.69.062320}.

\bibitem{ModeModeCoupling}
Johnson BR.
\newblock Controlling Photons in Superconducting Electrical Circuits.
\newblock Yale University; 2011.

\bibitem{PhysRevB.78.104508}
Mariantoni M, Deppe F, Marx A, Gross R, Wilhelm FK, Solano E.
\newblock Two-resonator circuit quantum electrodynamics: A superconducting
  quantum switch.
\newblock Phys Rev B. 2008 Sep;78:104508.
\newblock Available from:
  \url{http://link.aps.org/doi/10.1103/PhysRevB.78.104508}.

\bibitem{PurcellFiltering}
Reed MD, Johnson BR, Houck AA, DiCarlo L, Chow JM, Schuster DI, et~al.
\newblock Fast reset and suppressing spontaneous emission of a superconducting
  qubit.
\newblock Applied Physics Letters. 2010;96(20).
\newblock Available from:
  \url{http://scitation.aip.org/content/aip/journal/apl/96/20/10.1063/1.3435463}.

\bibitem{QuantumQuadratureMeasurements}
Wiseman HM, Milburn GJ.
\newblock Quantum theory of field-quadrature measurements.
\newblock Phys Rev A. 1993 Jan;47:642--662.
\newblock Available from:
  \url{http://link.aps.org/doi/10.1103/PhysRevA.47.642}.

\bibitem{NumericalSolutionSDEs}
Kloeden PE, Platen E.
\newblock Numerical Solution of Stochastic Differential Equations.
\newblock Stochastic Modelling and Applied Probability. Springer Berlin
  Heidelberg; 2013.
\newblock Available from: \url{https://books.google.de/books?id=r9r6CAAAQBAJ}.

\bibitem{PRepresentation}
Gambetta J, Blais A, Schuster DI, Wallraff A, Frunzio L, Majer J, et~al.
\newblock Qubit-photon interactions in a cavity: Measurement-induced dephasing
  and number splitting.
\newblock Phys Rev A. 2006 Oct;74:042318.
\newblock Available from:
  \url{http://link.aps.org/doi/10.1103/PhysRevA.74.042318}.

\bibitem{PointerStates}
Bonzom V, Bouzidi H, Degiovanni P.
\newblock Dissipative dynamics of circuit-QED in the mesoscopic regime.
\newblock The European Physical Journal D-Atomic, Molecular, Optical and Plasma
  Physics. 2008;47(1):133--149.

\bibitem{QuantumOptics}
Walls DF, Milburn GJ.
\newblock Quantum Optics.
\newblock Springer Study Edition. Springer Berlin Heidelberg; 2012.
\newblock Available from: \url{https://books.google.de/books?id=o6nrCAAAQBAJ}.

\bibitem{LTISystems}
Hespanha JP.
\newblock Linear Systems Theory.
\newblock Princeton University Press; 2009.
\newblock Available from: \url{https://books.google.de/books?id=tvd4ILdJUQoC}.

\bibitem{NetworkAnalysis}
Anderson BDO, Vongpanitlerd S.
\newblock Network Analysis and Synthesis: A Modern Systems Theory Approach.
\newblock Dover Books on Engineering. Dover Publications; 2013.
\newblock Available from: \url{https://books.google.de/books?id=MsbCAgAAQBAJ}.

\bibitem{ShermanMorrison}
Sherman J, Morrison WJ.
\newblock Adjustment of an Inverse Matrix Corresponding to a Change in One
  Element of a Given Matrix.
\newblock The Annals of Mathematical Statistics. 1950 Mar;21(1):124--127.
\newblock Available from: \url{http://www.jstor.org/stable/2236561}.

\bibitem{CoherentStates}
Glauber RJ.
\newblock Quantum Theory of Optical Coherence: Selected Papers and Lectures.
\newblock Wiley; 2007.
\newblock Available from: \url{https://books.google.de/books?id=9V3GzE6iqOYC}.

\bibitem{GenericLindbladForm}
Gorini V, Kossakowski A, Sudarshan ECG.
\newblock Completely positive dynamical semigroups of N-level systems.
\newblock Journal of Mathematical Physics. 1976;17(5):821--825.
\newblock Available from:
  \url{http://scitation.aip.org/content/aip/journal/jmp/17/5/10.1063/1.522979}.

\bibitem{BreuerPetruccione}
Breuer HP, Petruccione F.
\newblock The Theory of Open Quantum Systems.
\newblock Oxford University Press; 2002.
\newblock Available from: \url{https://books.google.de/books?id=0Yx5VzaMYm8C}.

\bibitem{HadamardProducts}
Havel TF, Sharf Y, Viola L, Cory DG.
\newblock Hadamard products of product operators and the design of
  gradient-diffusion experiments for simulating decoherence by NMR
  spectroscopy.
\newblock Physics Letters A. 2001;280(56):282 -- 288.
\newblock Available from:
  \url{http://www.sciencedirect.com/science/article/pii/S0375960101000834}.

\bibitem{NonMarkovianity}
Breuer HP, Laine EM, Piilo J.
\newblock Measure for the Degree of Non-Markovian Behavior of Quantum Processes
  in Open Systems.
\newblock Phys Rev Lett. 2009 Nov;103:210401.
\newblock Available from:
  \url{http://link.aps.org/doi/10.1103/PhysRevLett.103.210401}.

\bibitem{DetectionEstimationModulation}
Van~Trees HL.
\newblock Detection, Estimation, and Modulation Theory,.
\newblock Wiley; 2004.
\newblock Available from: \url{https://books.google.de/books?id=Xzp7VkuFqXYC}.

\bibitem{OptimalFiltering}
Gambetta J, Braff WA, Wallraff A, Girvin SM, Schoelkopf RJ.
\newblock Protocols for optimal readout of qubits using a continuous quantum
  nondemolition measurement.
\newblock Phys Rev A. 2007 Jul;76:012325.
\newblock Available from:
  \url{http://link.aps.org/doi/10.1103/PhysRevA.76.012325}.

\bibitem{FaultTolerantArchitectures}
{Fowler} AG, {Stephens} AM, {Groszkowski} P.
\newblock {High-threshold universal quantum computation on the surface code}.
\newblock Phys Rev A. 2009 Nov;80(5):052312.

\bibitem{steckNotes}
Steck DA.
\newblock Quantum and Atom Optics; 2007.
\newblock Available from: \url{steck.us/teaching}.

\bibitem{InPreparation}
Ciani A, Criger B, DiVincenzo D; 2016.
\newblock In preparation.

\end{thebibliography}



\end{backmatter}
\end{document}